\documentclass[a4paper]{article}       

\usepackage{amsmath,mathtools,amsthm,a4wide}
\usepackage{amssymb, authblk}
\usepackage[mathscr]{eucal}
\usepackage{bm}
\usepackage{bbm}
\usepackage[shortlabels]{enumitem}
\usepackage{hyperref}
\usepackage[usenames,dvipsnames]{xcolor}

\usepackage{tikz}
\usepackage{graphicx, caption,subfig}
\usepackage{todonotes}

\newcommand{\prob}{\mathbb{P}}
\newcommand{\Prob}[1]{\prob\left(#1\right)}

\newcommand{\expec}{\mathbb{E}}
\newcommand{\Exp}[1]{\expec\left[#1\right]}

\newcommand{\plim}{\ensuremath{\stackrel{\prob}{\longrightarrow}}}

\newcommand{\bigOp}[1]{O_{\sss\prob}\left(#1\right)}

\newcommand{\sss}[1]{\scriptscriptstyle{#1}}
\newcommand*{\swap}[2]{#2#1}

\newcommand{\me}{\textup{e}}
\newcommand{\dd}{{\rm d}}
\newcommand{\mean}[1]{\langle #1 \rangle}

\newcommand{\op}{o_{\sss\prob}}

\allowdisplaybreaks

\graphicspath{{Figures/}}

\begin{document}
\title{Closure coefficients in scale-free complex networks}

\author{Clara Stegehuis}
\affil{Twente University}

\maketitle

\begin{abstract}
	The formation of triangles in complex networks is an important network property that has received tremendous attention. The formation of triangles is often studied through the clustering coefficient. The closure coefficient or transitivity is another method to measure triadic closure. This statistic measures clustering from the head node of a triangle (instead of from the center node, as in the often studied clustering coefficient).
	We perform a first exploratory analysis of the behavior of the local closure coefficient in two random graph models that create simple networks with power-law degrees: the hidden-variable model and the hyperbolic random graph. 
		We show that the closure coefficient behaves significantly different in these simple random graph models than in the previously studied multigraph models.
		We also relate the closure coefficient of high-degree vertices to the clustering coefficient and the average nearest neighbor degree. 
\end{abstract}

\section{Introduction}

Networks describe the connectivity patterns of pairs of vertices. Examples of networks include social networks, the Internet, biological networks, the brain or communication networks. While these examples are very different in application, their connection patterns often share some characteristics. 
For example, in many real-world networks, vertices tend to cluster together in groups with relatively many edges between the group members~\cite{girvan2002} and networks often contain a large amount of triangles~\cite{ugander2011,watts1998}.

Usually the amount of triangles, or clustering, is measured in terms of the local clustering coefficient $c(k)$, the probability that two neighbors of a degree-$k$ vertex are neighbors themselves. In most real-world networks, the function $c(k)$ decreases in $k$ as some power law~\cite{vazquez2002,maslov2004,stegehuis2017,boguna2003,colomer2012,krioukov2012,stegehuis2019}. This indicates for example that two random friends of a popular person are less likely to know each other than two random friends of a less popular person. The local clustering curve of a network influences the spread of epidemic processes on a network, it contains information about its community structure, and is an important feature in classifying emails as spam or non spam, making it an important network property~\cite{serrano2006,colomer2013,becchetti2008}.

While the clustering coefficient turns out to be useful in some important network applications, other statistics that measure triadic closure may be more useful in other network applications. One promising such network statistic is the the closure coefficient~\cite{yin2019} or transitivity~\cite{frank1980,harary1979}. Where the clustering coefficient measures the fraction of times a vertex of degree $k$ serves as the center of a triangle, the closure coefficient measures the fraction of times a vertex of degree $k$ serves as the head of a triangle. Whereas the local clustering coefficient often decreases in $k$, the closure coefficient of many real-world networks is increasing instead~\cite{yin2019}, thus behaving substantially different from the local clustering coefficient. 
The closure coefficient was shown to be useful in link prediction~\cite{yin2019} and other types of network prediction tasks~\cite{yin2019a}. Furthermore, the conductance of the neighborhood of a vertex, a statistic that is often used in community detection, can be linked to its closure coefficient~\cite{yin2019}. Thus, the closure coefficient provides essential information about the tendency for network clustering which complements the information obtained from the clustering coefficient.

To be able to interpret the closure coefficient on real-world network data, it is crucial to understand its behavior in network null models, also known as random graph models. 
Whereas the behavior of the local clustering coefficient is well-understood in many types of random graphs~\cite{colomer2012,krioukov2012,stegehuis2019,krioukov2016,hofstad2017b}, the behavior of the closure coefficient is yet unknown in most random graph models. Only for one such model, the configuration model, the local closure coefficient was analyzed and was shown to be proportional to $k$~\cite{yin2019,yin2019a}. However, the configuration model creates multigraphs: graphs with self-loops and multiple edges, whereas most real-world networks are simple networks. This makes the configuration model unfit for understanding real-world networks. Therefore, analyzing the behavior of the closure coefficient in random graph models that create simple networks instead is an important problem. 
However, imposing simplicity constraints on a random graph model often makes it more difficult to analyze, because of the non-trivial degree-degree correlations that arise from such simplicity constraints~\cite{hofstad2017,catanzaro2005,yao2017}.

Degree-correlations often become more pronounced when the degree distribution is scale-free.
Scale-free networks describe network connections with strong degree heterogeneity, often modeled as a power law where the proportion of vertices with $k$ neighbors scales as $k^{-\tau}$. In many real-world networks, the power-law exponent $\tau$ was found to lie between 2 and 3 \cite{albert1999,faloutsos1999,jeong2000,vazquez2002}, so that the vertex degrees have a finite first and infinite second moment. These power laws make vertices of extremely high degrees (also called hubs) likely to be present, causing non-trivial degree-correlations~\cite{ostilli2014,hofstad2017}.

In this paper, we focus on the closure coefficient in two substantially different simple network models that create power-law degrees.
We first investigate the closure coefficient in the hidden-variable model. In the absence of high-degree vertices, this model creates graphs that are similar to the configuration model~\cite{hofstad2009}. With power-law degrees however, the degree-degree correlations make these two models significantly different~\cite{hofstad2017}. We show that this results in considerably different behavior for the local closure coefficient in the hidden-variable model than in the configuration model. 

The configuration model, as well as the hidden-variable model do not contain many triangles in the large-network limit. Real-world networks on the other hand, often contain high levels of clustering~\cite{girvan2002}. 
We therefore also investigate the closure coefficient in the hyperbolic model, which in recent years has emerged as an important scale-free network model \cite{krioukov2010,allard2017,boguna2010,garcia-perez2016,borassi2015}. The hyperbolic model creates networks that simultaneously possess two crucial characteristics of real-world networks: power-law degrees and clustering. Therefore, the behavior of the closure coefficient in the hyperbolic model could potentially serve as a benchmark for the behavior of the closure coefficient in real-world network data.

We then relate the closure coefficient to two other frequently used network statistics: the local clustering coefficient and the average nearest neighbor degree of a vertex. 
Specifically, we show that the closure coefficient of a vertex can be related to a ratio of its clustering coefficient with its average nearest neighbor degree. Since the behavior of these statistics is known for many random graph models, this provides a simple method to investigate the closure coefficient of many network models.

The average local closure coefficient over all degree $k$ vertices is difficult to analyze, as it averages over a ratio of two correlated random variables. We therefore introduce the `typical' closure coefficient, a statistic that enables to investigate the behavior of the closure coefficient of large-degree vertices. In the large-network limit, this typical closure coefficient behaves similarly to the median value of the closure coefficient over all degree-$k$ vertices. 

We also analyze the higher-order closure coefficient, measuring the closure of cliques of size $l$ into larger cliques of size $l+1$. These higher order closure coefficients bound the motif conductance of the neighborhood of a vertex, which is useful in community detection~\cite{yin2019}. 
 In this paper, we provide a first exploratory analysis of the asymptotic behavior of these higher-order closure coefficients for both the hyperbolic model and the hidden-variable model. We obtain these asymptotics by optimizing the structure of network cliques, a method which has recently been proposed to analyze subgraphs and local clustering~\cite{stegehuis2019,stegehuis2019b}.



\paragraph*{Overview of the paper.}
We first present the definitions of the local closure coefficient as well as the higher-order closure coefficients in Section~\ref{sec:closuredef}. We then analyze the behavior of the closure coefficient in the hidden-variable model in Section~\ref{sec:hid} and for the hyperbolic random graph in Section~\ref{sec:hyp}.

\paragraph*{Notation}
We write $\plim $ for convergence in probability. We say that a sequence of events $(\mathcal{E}_n)_{n\geq 1}$ happens with high probability if $\lim_{n\to\infty}\Prob{\mathcal{E}_n}=1$. Furthermore, we write $X_n=\bigOp{g(n)}$ for a sequence of random variables $(X_n)_{n\geq1}$ when $|X_n|/g(n)$ is a tight sequence of random variables and $X_n=o_{\sss{\prob}}(g(n))$ if $X_n/g(n)\plim 0$. We say that $X_n\propto_{\sss{\prob}} g(n)$ if for any $h(n)$ such that $\lim_{n\to\infty}h(n)=\infty$,
\begin{equation}
\lim_{n\to\infty}\Prob{\frac{X_n}{g(n)}<\frac{1}{h(n)}}= 0 \quad \text{and}\quad  \lim_{n\to\infty}\Prob{\frac{X_n}{g(n)}>h(n)}= 0. 
\end{equation}
In particular, this means that when $X_n/g(n)\plim c$ for some constant $c$, then also $X_n\propto_{\sss{\prob}} g(n)$. 
Furthermore, $X_n=\bigOp{g(n)}$ implies that $X_n\propto_{\sss{\prob}} g(n)$.
Finally, we denote $[k]=1,2,\dots,k$. 

\section{Closure coefficients}\label{sec:closuredef}
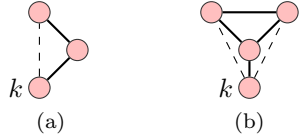
\begin{figure}[tb]
	\tikzstyle{every node}=[circle,fill=red!25,minimum size=8pt,inner sep=0pt,draw=black!80]
	\centering
	\subfloat[]{
	\begin{tikzpicture}
	\centering
	\tikzstyle{edge} = [draw,thick,-]
	\node[label=left:{$k$}] (a) at (0,0) {};
	\node[] (c) at (0,1) {};
	\node[] (e) at (0.5,0.5) {};
	\draw[dashed] (a)--(c);
	\draw[edge] (e)--(c);
	\draw[edge] (a)--(e);
	\end{tikzpicture}
	\label{fig:closure}
}
\hspace{1cm}
\subfloat[]{
	\begin{tikzpicture}
	\centering
	\tikzstyle{edge} = [draw,thick,-]
	\node[label=left:{$k$}] (a) at (0.5,0) {};
	\node[] (c) at (0,1) {};
	\node[] (e) at (0.5,0.5) {};
	\node[] (b) at (1,1) {};
	\draw[dashed] (a)--(c);
	\draw[dashed] (a)--(b);
	\draw[edge] (e)--(c);
	\draw[edge] (a)--(e);
	\draw[edge] (b)--(c);
	\draw[edge] (b)--(e);
	\end{tikzpicture}
	\label{fig:chigh}
}
\caption{The closure coefficient investigates the probability that a two-path starting from a degree-$k$ vertex closes into a triangle (with the dashed edge), illustrated in (a). (b) illustrates $H_3(k)$. The solid lines from a 3-wedge, and $H_3(k)$ computes the probability that the dashed lines are present.}
\end{figure}

The closure coefficient, or transitivity of vertex $v$, $H(v)$, is defined by~\cite{frank1980,harary1979,yin2019}
\begin{equation}
H(v) = \frac{2\triangle_v}{W(v)},
\end{equation}
where $\triangle_v$ denotes the number of triangles attached to vertex $v$ and $W(v)$ the number of two-paths from vertex $v$. The coefficient 2 accounts for the fact that each triangle contains two different length-2 paths from node $v$. Thus, the closure coefficient measures the fraction of two-paths that merge into triangles, as illustrated in Fig.~\ref{fig:closure}. The average local closure coefficient is then defined as the average of this quantity over all  vertices of degree $k$,
\begin{equation}\label{eq:avgclosure}
H^{(a)}(k)= \frac{1}{N_k}\sum_{i:d_i=k}\frac{\triangle_{i}}{W(i)},
\end{equation}
where $N_k$ denotes the number of vertices of degree $k$, and $d_i$ denotes the degree of vertex $i$. 

\begin{figure*}[tbp]
	\centering
	\subfloat[]{
		\centering
		\includegraphics[width=0.4\textwidth]{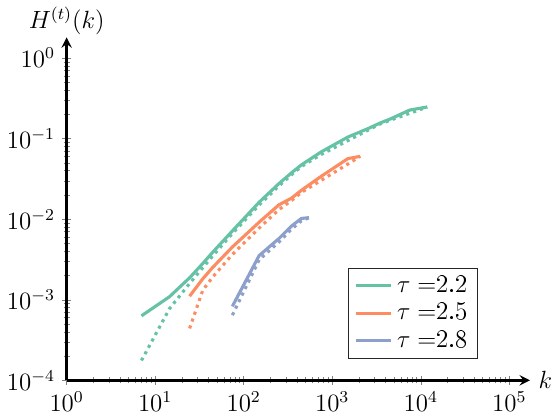}
		\label{fig:clakhvm}
	}
	\hspace{0.4cm}
	\subfloat[]{
		\centering
		\includegraphics[width=0.4\textwidth]{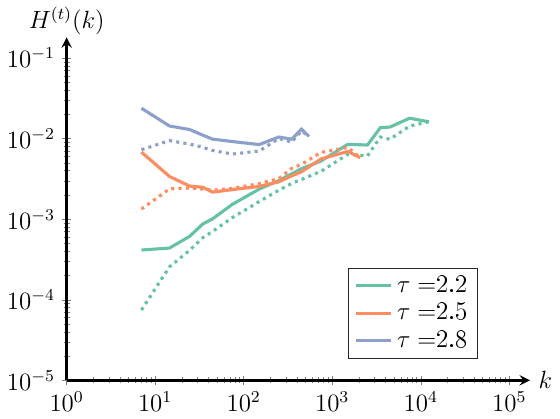}
		\label{fig:clakhyp}
	}
	\caption{The solid line plots $k c(k)/(2a(k))$ whereas the dashed line plots the median closure coefficient averaged over $10^4$ network realizations, for $n=10^5$ and various values of $\tau$ for a) the hidden variable model b) the hyperbolic random graph.}
	\label{fig:clak}
\end{figure*}

\subsection{Higher-order closure}
Where the closure coefficient measures the fraction of two-paths merging into triangles, higher order closure coefficients measure the tendency of forming larger cliques.

 We define an $l$-wedge from vertex $v$ as an edge incident to vertex $v$ connecting to some other vertex $u$, such that $u$ is part of an $l$-clique~\cite{yin2018}. See Fig.~\ref{fig:chigh} for an example. Let $W^{(l)}(v)$ denote the number of $l$-wedges attached to vertex $v$. Furthermore, let $K_l(v)$ denote the number of cliques incident to vertex $v$. Then, the $l$-th order closure coefficient is defined as the fraction of $l$-wedges that close into $l+1$ cliques containing vertex $v$. Formally,
\begin{equation}\label{eq:clhigh}
H_l(v) = \frac{lK_{l+1}(v)}{W^{(l)}(v)},
\end{equation}
where the coefficient $l$ accounts for the fact that each $l+1$ clique contains $l$ different $l$-wedges.
Then $H_l^{(a)}(k)$ is again defined as the average of this quantity over all vertices of degree $k$.

\subsection{Typical closure}\label{sec:typ}
The behavior of the numerator as well as the denominator of the local closure coefficient can be analyzed easily. However, when taking the average of these random ratios over all vertices of degree $k$, some rare values with a large ratio may dominate the value of the average local closure coefficient. Furthermore, the average over these ratios of correlated random variables is difficult to analyze mathematically. In this paper, we therefore focus on the `typical' closure coefficient of vertices with degree $k$ instead of its average. Typical closure can be interpreted as the behavior of the closure coefficient of a randomly chosen vertex of degree $k$ and we denote the typical closure coefficient of a vertex of degree $k$ by $H^{(t)}(k)$.
Formally, let $V_k$ denote a randomly chosen vertex of degree $k$. Then, we define the typical closure coefficient as
\begin{equation}\label{eq:closuretypicaldef}
H^{(t)}(k) =\frac{\triangle_{V_k}}{W(V_k)}.
\end{equation}
Thus, the typical closure coefficient divides the number of triangles adjacent to a randomly chosen vertex by the number of two-paths from the \emph{same} vertex.
This typical closure is more informative than average closure, since it provides information on the closure of a randomly chosen vertex of degree $k$, instead of on the closure of vertices of degree $k$ with extremely high closure values that may dominate the average. Furthermore, the typical closure coefficient is easier to analyze, as it allows to analyze the numerator and denominator separately. Similarly, we denote the typical higher-order closure coefficient by $H_l^{(t)}(k)$, which is defined as the higher-order closure coefficient of a randomly chosen vertex of degree $k$.

In the rest of this paper, we will provide an exploratory analysis of the asymptotic behavior of these clustering coefficients. That is, we investigate the behavior of $H^{(t)}(k_n)$ as the network size $n$ tends to infinity, while $k_n$ grows as a function of $n$.

\paragraph{Relation with median closure.}
For a given network data set, the typical closure coefficient is a random variable, which is problematic when computing a statistic of a network data set. However, in this paper, we show that in the large-network limit, the rescaled typical closure coefficient concentrates in the large-network limit. This means that all but a vanishing fraction of degree $k$ vertices satisfy the predicted scaling, so that the median closure coefficient follows the same scaling. Therefore, a proxy for the typical closure coefficient of a given network is the median of the typical closure coefficient. The median closure coefficient takes the median value of $\triangle_{i}/W(i)$ over all vertices of degree $k$, which is a deterministic statistic. The higher-order median closure coefficients are defined similarly.
In this paper, we will therefore compute the median closure coefficient in our simulations, and compare them with our mathematical analysis of the typical closure coefficient. 

\subsection{Relation with $a(k)$ and $c(k)$}\label{sec:ckak}
We now argue that in many random graph models, the typical closure coefficient can be related to two other important network statistics: the average nearest neighbor degree $a(k)$ and the local clustering coefficient $c(k)$. 
The average nearest neighbor degree $a(k)$ is defined as
\begin{equation}
a(k) = \frac{1}{k N_k}\sum_{i:d_i=k}\sum_{j\in \mathcal{N}_i}d_j,
\end{equation}
where $\mathcal{N}_i$ denotes the set of neighbors of vertex $i$ and $N_k$ again denotes the number of vertices of degree $k$.

The local clustering coefficient $c(k)$ measures the fraction of pairs of neighbors of a vertex that close into triangles. Thus, the local clustering coefficient measures a similar tendency as the local closure coefficient, with the difference that the clustering coefficient measures the fraction of triangles that are formed with $v$ as the center of the triangle, whereas the closure coefficient measures the fraction of triangles with $v$ as the head of the triangle. Formally, the local clustering coefficient of vertex $v$ is defined as
\begin{equation}
c(v) = \frac{\triangle_v}{d_v(d_v-1)/2}.
\end{equation}
Usually, the local clustering coefficient is analyzed by averaging over all vertices of degree $k$, which is denoted by $c(k)$. A notable difference between $c(k)$ and the local closure coefficient is that whereas $c(k)$ has a random numerator (the number of triangles at a vertex of degree $k$) but a deterministic denominator ($k(k-1)/2$), the closure coefficient has both a random numerator and a random denominator, making the analysis of the closure coefficient more involved. 

We now explain the relation between $H^{(t)}(k)$ and the two other network statistics $a(k)$ and $c(k)$. 
The average number of wedges from a vertex of degree $k$ equals $k(a(k)-1)$ (where the $-1$ accounts for the fact that one of the connections of a neighbor is used to connect to the degree $k$ vertex itself). Also, the average number of triangles attached to a vertex of degree $k$ equals $k^2c(k)/2$. 

Furthermore, for $k$ sufficiently large, the number of wedges from a vertex of degree $k$ concentrates around $k(a(k)-1)\approx ka(k)$ in many random graph models~\cite[Proposition~2]{stegehuis2017b}. That is, if $W(V_k)$ denotes the number of wedges from a randomly chosen vertex of degree $k$, then in many random graph models
\begin{equation}
\frac{W(V_k)}{k a(k)}\plim 1. 
\end{equation}

Similarly, for $k$ sufficiently large, the number of triangles attached to a vertex of degree $k$ concentrates around $k^2c(k)/2$ for several random graph models~\cite{stegehuis2019}. Thus, if $\triangle_{V_k}$ denotes the number of triangles attached to a randomly chosen vertex of degree $k$, then in many random graph models
\begin{equation}
\frac{\triangle_{V_k}}{k^2c(k)/2}\plim 1.
\end{equation}
This yields for the typical closure coefficient that
\begin{equation}\label{eq:closakck}
\frac{H^{(t)}(k)}{{kc(k)}/{(2a(k))}}\plim 1.
\end{equation}
For many random graph models $c(k)$ and $a(k)$ are known. Furthermore, for $k$ sufficiently large, the number of wedges and triangles concentrate around these values for many random graph models. Therefore, this gives an easy method to investigate the local closure coefficient of many types of random graph models when $k$ is sufficiently large. 

Figure~\ref{fig:clak} illustrates the relationship between $H^{(t)}(k)$, $a(k)$ and $c(k)$ for two different random graph models. For the hidden-variable model (which will be introduced in Section~\ref{sec:hid})),~\eqref{eq:closakck} also holds for small values of $k$, whereas for the hyperbolic model (which will be introduced in Section~\ref{sec:hyp})), the fit only holds for larger values of $k$.

%
%
%

\section{Hidden variable model}\label{sec:hid}
The hidden-variable model creates a random graph on $n$ vertices. 
Every vertex $i$ is equipped with a weight, $h_i$. These weights are drawn independently from the distribution
\begin{equation}\label{eq:deg_distr}
\rho(h) = Ch^{-\tau},
\end{equation}
for some $\tau\in(2,3)$, so that the largest weight $\max_i h_i=\bigOp{n^{1/(\tau-1)}}$. Every pair of vertices $(h, h')$ is then connected with probability $p(h, h')$. In this paper we work with (although many
other choices are possible)
\begin{equation}\label{eq:conprobhvm}
p(h,h') = \min\Big(\frac{hh'}{\mean{h}n},1\Big),
\end{equation}
which is equivalent to the Chung-Lu model~\cite{chung2002}. Here $\mean{h}$ denotes the expected vertex weight. 
In the large-network limit, the degree of vertex $i$ will be close to its hidden variable $h_i$. Thus, the hidden-variable model creates simple, power-law random graphs in this manner where the maximal degree scales as $n^{1/(\tau-1)}$.

\subsection{Closure coefficient}
We now investigate the closure coefficient in the hidden-variable model using the argument of Section~\ref{sec:ckak}. 
The number of wedges attached to a vertex of degree $k$ with $k\gg n^{(\tau-2)/(\tau-1)}$ scales as $ka(k)$ with high probability~\cite{stegehuis2017b} in the hidden variable model. Furthermore, for $ n^{(\tau-2)/(\tau-1)}\ll k\ll n^{1/(\tau-1)}/\log(n)$~\cite{stegehuis2017b},
\begin{equation}
\frac{a(k)}{n^{3-\tau}k^{\tau-3}} \plim \frac{C\mean{h}^{2-\tau}}{(3-\tau)(\tau-2)}.
\end{equation}

Similarly, by Appendix~\ref{sec:ckhvm},
\begin{equation}\label{eq:ckhvm}
c(k) =  \begin{cases}
n^{2-\tau}\log(n/k^2)\frac{C^2\mean{h}^{-\tau}}{(3-\tau)(\tau-2)}(1+\op(1)) &  n^{(\tau-2)/(\tau-1)}\ll k \ll \sqrt{n}\\
n^{5-2\tau}k^{2\tau-6}\frac{C^2\mean{h}^{3-2\tau}}{(3-\tau)^2(\tau-2)^2}(1+\op(1)) & \sqrt{n}\ll k\ll n^{1/(\tau-1)}/\log(n).
\end{cases}
\end{equation}

Thus, by~\eqref{eq:closakck}, for $k\gg n^{(\tau-2)/(\tau-1)}$, the typical closure coefficient of a vertex of degree $k$ in the hidden-variable model satisfies
\begin{equation}\label{eq:Hkhvm}
H^{(t)}(k)= \begin{cases}
n^{-1}k^{4-\tau}\log(n/k^2)C\mean{h}^{-2}(1+\op(1)) & n^{(\tau-2)/(\tau-1)}\ll k \ll \sqrt{n}\\
n^{2-\tau}k^{\tau-2}\frac{\mean{h}^{1-\tau}C}{2(3-\tau)(\tau-2)}(1+\op(1)) &  \sqrt{n}\ll k \ll n^{1/(\tau-1)}/\log(n).
\end{cases}
\end{equation}

Note that this behavior is significantly different from the $H^{(a)}(k)\propto k$ behavior found in the configuration model~\cite{yin2019}. This difference is caused by the numerous multiple edges appearing in the configuration model for $\tau\in(2,3)$, as well as by the difference between the behavior of the average closure coefficient versus its typical behavior.

\begin{figure}[tb]
	\centering
	\includegraphics[width=0.4\textwidth]{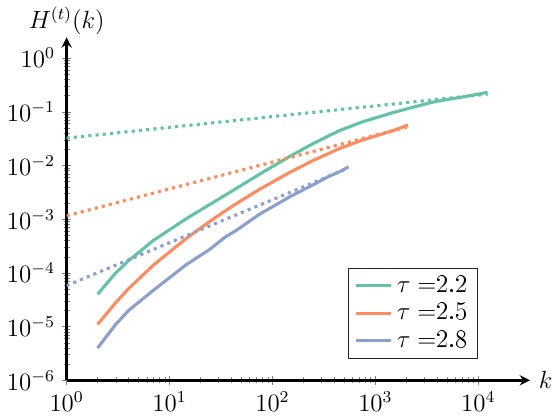}
	\caption{The median value of the closure coefficient averaged over $10^4$ realizations of hidden-variable models with $n=10^6$ and various values of $\tau$. The dashed line gives the asymptotic slope for $k\gg\sqrt{n}$ predicted by Eq.~\eqref{eq:Hkhvm}.}
	\label{fig:clashvm}
\end{figure}

\begin{figure*}[tb]
	\centering
	\subfloat[]{
		\centering
		\includegraphics[width=0.45\textwidth]{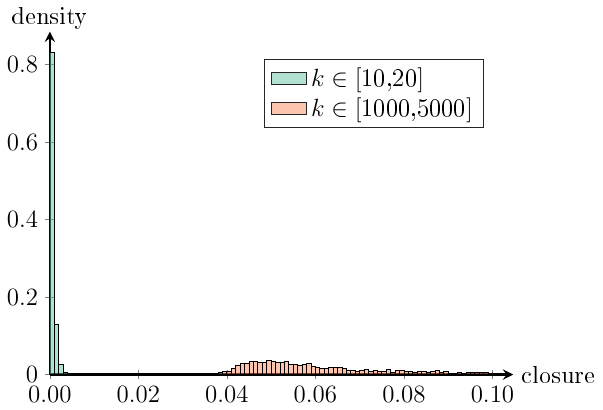}
		\label{fig:histhvm}
	}
	\hspace{0.4cm}
	\subfloat[]{
		\centering
		\includegraphics[width=0.45\textwidth]{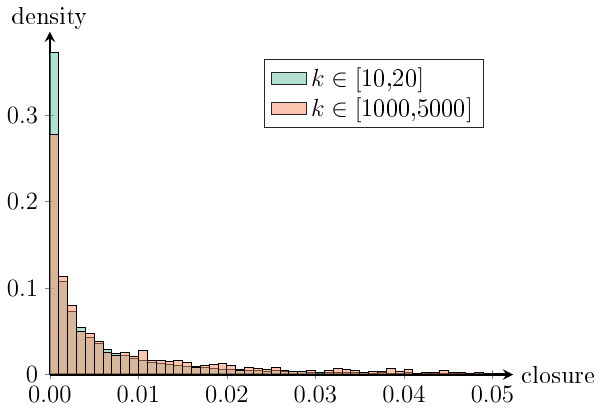}
		\label{fig:hishyp}
	}
	\caption{Density plot of closure coefficients for two ranges of $k$ and $n=10^6$, $\tau=2.5$ in a) the hidden-variable model and b) the hyperbolic model.}
	\label{fig:histclosure}
\end{figure*}

\paragraph{Relation with median closure.}
As~\eqref{eq:Hkhvm} states that the rescaled local closure coefficient converges in probability to a constant, this means that all but a vanishing fraction of vertices of degree $k$ has local closure coefficient close to~\eqref{eq:clhid}. Therefore, the typical local closure coefficient behaves the same as the median of the local closure coefficient over all vertices of degree $k$.
Figure~\ref{fig:clashvm} shows that Eq.~\eqref{eq:Hkhvm} indeed describes the asymptotic behavior of the median closure coefficient well. Figure~\ref{fig:histhvm} shows that indeed the distribution of the closure coefficient is skewed for small values of $k$, whereas it becomes less skewed for larger values of $k$. This implies that the difference between the typical and the average closure coefficient decreases as $k$ gets larger.

\subsection{Higher-order closure}
In this section, we provide an exploratory analysis of the the typical higher-order closure coefficient of a vertex of degree $k$ in the hidden variable model. 
To obtain this closure coefficient, it is convenient to study the typical higher-order closure coefficient of a vertex of \emph{weight} $h$ instead, which we denote by $\bar{H}^{(t)}_l(h)$. That is, when $V_h$ denotes a randomly chosen vertex of weight $h$, then 
\begin{equation}\label{eq:clhighweight}
\bar{H}^{(t)}_l(h) = \frac{lK_{l+1}(V_h)}{W^{(l)}(V_h)}.
\end{equation}
This is convenient, as the connection probabilities in Eq.~\eqref{eq:conprobhvm} are defined in terms of weights as well. In Appendix~\ref{sec:weightdegree} we then show that because the degree of a vertex is tighly concentrated around its weight, the higher-order closure coefficient of a vertex of given weight is close to the typical higher-order closure coefficient of a vertex of \emph{degree} $k$, $H^{(t)}_l(k)$. 

\paragraph{Cliques attached to a vertex of weight $h$.}
First, we analyze the numerator of Eq.~\eqref{eq:clhighweight}. Thus, we investigate the typical number of cliques of size $l+1$ attached to a vertex of weight $h$.  
To obtain this, we find the optimal clique structure in terms of the weights of vertices involved in the clique as in~\cite{stegehuis2019b,stegehuis2019}. The probability that the vertex of weight $h$ attaches to a clique of size $l$ can be written as
\begin{equation}\label{eq:cliquecontr}
\Prob{\text{clique}}=\int_{\boldsymbol{h}}\Prob{\text{clique with vertices of weights }\boldsymbol{h}}\rho(h_1)\rho(h_2)\dots\rho(h_l)\dd\boldsymbol{h}
\end{equation}
where the integral is over all possible weight sequences $\boldsymbol{h}=(h_1,\dots,h_l)$ of the $l$ vertices involved in the clique. We now let $h_1,\dots,h_l$ scale as $n^{\alpha_1},\dots,n^{\alpha_l}$ and find which weights give the largest contribution to the integrand in~\eqref{eq:cliquecontr}.

We compute the probability that a randomly chosen vertex of weight $h$ forms a clique with $l$ neighbors with weights proportional to $(n^{\alpha_i})_{i\in[l]}$. By~\eqref{eq:conprobhvm}, the probability that these vertices form a clique with the weight-$h$ vertex is proportional to
\begin{align}
&\Prob{\text{clique of vertices of weights }n^{\alpha_i}\text{ with }V_h} \propto \prod_{i<j}\min(n^{\alpha_1+\alpha_j-1},1)\prod_{j\in[l]}\min(hn^{\alpha_j-1},1),
\end{align}
where the second product denotes the probability that all vertices connect to the weight-$h$ vertex, and the first product denotes the probability that all other vertices connect. 
By~\eqref{eq:deg_distr}, with high probability there are proportionally $n^{1+\alpha_i(1-\tau)}$ vertices of weight proportional to $n^{\alpha_i}$. 
Thus, the number of cliques containing the weight-$h$ vertex and $l$ vertices of weight $(n^{\alpha_i})_{i\in[l]}$ scales as
\begin{align}\label{eq:cliquekhid}
 \text{ \# cliques with }&V_h\text{ and vertices of weights }(n^{\alpha_i})_{i\in[l]}\nonumber\\
&\propto_{\sss{\prob}} n^{l+\sum_i\alpha_i(1-\tau)}\prod_{j\in[l]}\min(hn^{\alpha_j-1},1)\prod_{i<j}\min(n^{\alpha_1+\alpha_j-1},1).
\end{align}
Similarly as in~\cite[Theorem 2.1]{hofstad2017d}, we can show that for $h\gg 1$,
\begin{equation}\label{eq:cliquemaxeq}
K_{l+1}(V_h)\propto_{\sss{\prob}}\max_{\alpha_1,\dots,\alpha_l\in[0,1/(\tau-1)]}n^{l+\sum_i\alpha_i(1-\tau)}\prod_{j\in[l]}\min(hn^{\alpha_j-1},1)\prod_{i<j}\min(n^{\alpha_1+\alpha_j-1},1),
\end{equation}
if this equation has a unique maximizer over $\alpha_1,\dots,\alpha_l\in[0,1/(\tau-1)]$. Here the constraint $\alpha\in[0,1/(\tau-1)]$ comes from the fact that the maximal weight sampled from~\eqref{eq:deg_distr} is proportional to $n^{1/(\tau-1)}$ with high probability. 

For $l\geq 3$~\eqref{eq:cliquekhid} is uniquely maximized at $\alpha_i=1/2$ for all $i$. Plugging this optimizer into~\eqref{eq:cliquemaxeq} shows that the number of complete graphs of size $l+1$ attached to a typical vertex of weight $h$ scales as
\begin{equation}\label{eq:Klhid}
K_{l+1}(V_h)\propto_{\sss{\prob}} n^{l(3-\tau)/2}\min(hn^{-1/2},1)^l.
\end{equation}
Optimizing~\eqref{eq:cliquekhid} not only gives the scaling of the number of cliques, it also gives the structure of the most likely clique attached to a weight-$h$ vertex. That is, the fact that~\eqref{eq:cliquekhid} is optimized by $\alpha_i=1/2$ for all $i$ shows that almost all cliques attached to vertices of weight $h$ are cliques where the weights of the other vertices involved are proportional to $\sqrt{n}$.

\paragraph{Wedges attached to a vertex of weight $h$.}
We now analyze the denominator of~\eqref{eq:clhigh} and compute the typical number of $l$-wedges from a vertex of weight $h$. That is, we count the number of edges from the weight-$h$ vertex attached to a complete graph of size $l$. We use a similar optimization problem as for the number of $l$-cliques and now compute the probability that a vertex of weight $h$ is attached to a clique with $l$ vertices with weights proportional to $(n^{\alpha_i})_{i\in[l]}$. W.l.o.g. we assume that the weight-$h$ vertex attaches to vertex 1 in the clique (the vertex with weight proportional to $n^{\alpha_1}$). By~\eqref{eq:conprobhvm}, the probability that the weight-$h$ vertex connects to vertex 1 and that vertex 1 forms a clique with the other $l-1$ vertices scales as
\begin{align}
&\Prob{\text{weight-}h \text{ vertex attached to $l$-clique with weights }n^{\alpha_i}}\nonumber\\
& \propto \min(hn^{\alpha_1-1},1)\prod_{1\leq i<j\leq l}\min(n^{\alpha_1+\alpha_j-1},1).
\end{align}
Here the first term denotes the probability that the weight-$h$ vertex connects to the vertex of weight $n^{\alpha_1}$, and the second term denotes the probability that vertex 1 forms a clique with the other $l-1$ vertices. 
Again, by~\eqref{eq:deg_distr}, with high probability, there are $n^{1+\alpha_i(1-\tau)}$ vertices of weight proportional to $n^{\alpha_i}$. Thus, the number of $l$-wedges attached to the weight-$h$ vertex where the member of the $l$-wedge have weights $(n^{\alpha_i})_{i\in[l]}$ satisfies 
\begin{align}\label{eq:wedgehvmopt}
\text{ \# $l$-wedges }&V_h\text{ with vertices of weights }(n^{\alpha_i})_{i\in[l]}\nonumber\\
&\propto_{\sss{\prob}}n^{l+\sum_i\alpha_i(1-\tau)}\min(hn^{\alpha_1-1},1)\prod_{i<j}\min(n^{\alpha_1+\alpha_j-1},1).
\end{align}
Again, similarly to~\eqref{eq:cliquemaxeq}, we can show that for $h\gg 1$,
\begin{equation}\label{eq:wedgemaxeq}
W^{(l)}(V_h)\propto_{\sss{\prob}}\max_{\alpha_1,\dots,\alpha_l\in[0,1/(\tau-1)]}n^{l+\sum_i\alpha_i(1-\tau)}\min(hn^{\alpha_1-1},1)\prod_{i<j}\min(n^{\alpha_1+\alpha_j-1},1),
\end{equation}
if this equation has a unique maximizer over $\alpha_1,\dots,\alpha_l\in[0,1/(\tau-1)]$. For $l\geq 3$, this equation is indeed uniquely maximized for ${\alpha_i}=1/2$ for all $i$. 
Plugging this optimizer into~\eqref{eq:wedgemaxeq} yields that the typical number of $l$-wedges attached to a vertex of weight $h$ scales as
\begin{equation}
W^{(l)}(V_h)\propto_{\sss{\prob}} n^{l(3-\tau)/2}\min(hn^{-1/2},1).
\end{equation}
Combining this with~\eqref{eq:Klhid} gives for $l\geq 3$,
\begin{equation}\label{eq:chighhidweight}
\bar{H}_l^{(t)}(h)\propto_{\sss{\prob}} \begin{cases}
h^{l-1}n^{-(l-1)/2} & 1\ll h\ll \sqrt{n}\\
1 & h\gg\sqrt{n}.
\end{cases}
\end{equation}
In Appendix~\ref{sec:weightdegree} we show that this implies that
\begin{equation}\label{eq:chighhid}
{H}_l^{(t)}(k)\propto_{\sss{\prob}} \begin{cases}
k^{l-1}n^{-(l-1)/2} & 1\ll k\ll \sqrt{n}\\
1 & k\gg\sqrt{n}.
\end{cases}
\end{equation}

Interestingly, the closure coefficient becomes independent of $k$ for $k\gg\sqrt{n}$. This is caused by the core of the hidden-variable model, where most vertices of degrees $\sqrt{n}$ and higher form a giant clique. A clique in this core is very likely to form a larger clique with additional vertices of degree $\sqrt{n}$ and higher, explaining the constant higher-order closure coefficient of these vertices. 

\begin{figure*}[tb]
	\centering
	\subfloat[]{
		\centering
		\includegraphics[width=0.4\textwidth]{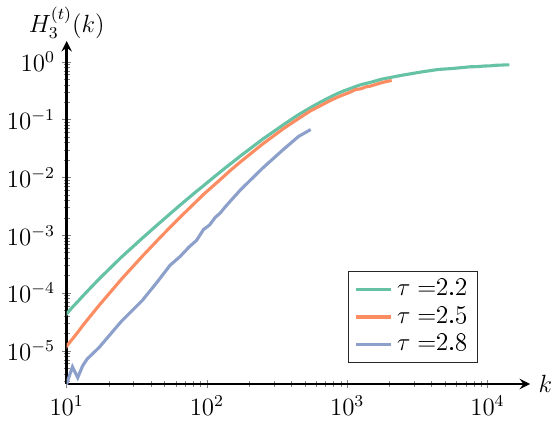}
		\label{fig:cl3hvm}
	}
	\hspace{0.4cm}
	\subfloat[]{
		\centering
		\includegraphics[width=0.4\textwidth]{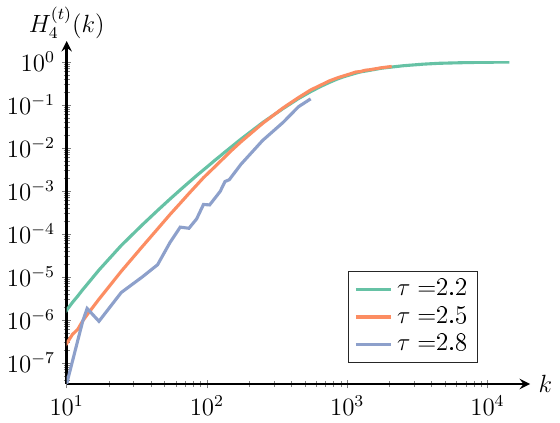}
		\label{fig:cl4hvm}
	}
	\caption{The median value of higher-order closure coefficients in the hidden-variable model with $n=10^6$ and various values of $\tau$. a) $H_3^{(t)}(k)$, b) $H_4^{(t)}(k)$.}
	\label{fig:clhighhvm}
\end{figure*}

\paragraph{Relation with median closure.}
As the typical closure coefficient scales as~\eqref{eq:chighhid} with high probability, this suggests that the large-network behavior of the typical closure coefficient is similar to the median higher-order closure coefficient. 
Figure~\ref{fig:clhighhvm} shows the behavior of the median higher-order closure coefficients in the hidden-variable model. Indeed, when $k$ gets sufficiently large, these median higher-order closure coefficients approach a constant value, as predicted by Eq.~\ref{eq:chighhid}. Note that for $\tau$ close to 3, this is less pronounced, as the largest degree in a power-law network scales as $n^{1/(\tau-1)}$, which is close to $\sqrt{n}$.

\section{Hyperbolic model}\label{sec:hyp}
The hyperbolic random graph~\cite{krioukov2010} is a key model that creates simple power-law random graphs with non-trivial clustering by exploiting an underlying geometric space. 
The hyperbolic random graph samples $n$ vertices in a disk of radius $R=2\log(n/\nu)$, where $\nu$ is a parameter that influences the average degree of the generated networks. The density of the radial coordinate $r$ of a vertex $p=(r,\phi)$ is
\begin{equation}
\rho(r)=\beta\frac{\sinh(\beta r)}{\cosh(\beta R)-1}
\end{equation}
with $\beta=(\tau-1)/2$. The angle $\phi$ of $p$ is sampled uniformly from $[0,2\pi]$. Two vertices are connected if their hyperbolic distance is at most $R$. The hyperbolic distance of points $u=(r_u,\phi_u)$ and $v=(r_v,\phi_v)$ satisfies
\begin{align}
\cosh(\dd (u,v))=& \cosh(r_u)\cosh(r_v) -\sinh(r_u)\sinh(r_v)\cos(\Delta\theta),
\end{align}
where $\Delta\theta$ denotes the angle between $u$ and $v$.
Two neighbors of a vertex are likely to be close to one another due to the geometric nature of the hyperbolic random graph. Therefore, the hyperbolic random graph contains many triangles~\cite{gugelmann2012}. Furthermore, the model generates scale-free networks with degree exponent $\tau$~\cite{krioukov2010}. 

\subsection{Closure coefficient}
We now investigate the typical closure coefficient of a vertex in the hyperbolic random graph using the analysis in Section~\ref{sec:ckak}. 
Similarly as in the hidden-variable model~\cite{stegehuis2017b}, for $n^{(\tau-2)/(\tau-1)}\ll k\ll n^{1/(\tau-1)}/\log(n)$,
\begin{equation}
a(k) \propto_{\sss{\prob}} 
n^{3-\tau}k^{\tau-3}.
\end{equation}
Moreover, for $k\gg n^{(\tau-2)/(\tau-1)}$ the number of wedges attached to a vertex of degree $k$ concentrates around $ka(k)$~\cite{stegehuis2017b}.
 Furthermore, in the hyperbolic model
 \begin{equation}
 c(k)\propto_{\sss{\prob}} \begin{cases}
 k^{-1} & \tau>5/2\\
 k^{4-2\tau} & \tau<5/2, k\ll \sqrt{n}\\
 k^{2\tau-6}n^{5-2\tau}& \tau<5/2, k\gg\sqrt{n},
 \end{cases}
 \end{equation}
 and the number of triangles concentrates around $k^2c(k)$ for $k\gg 1$~\cite{stegehuis2019}. 
 
 This yields 
 \begin{equation}\label{eq:clhid}
 H^{(t)}(k)\propto_{\sss{\prob}} \begin{cases}
 n^{\tau-3}k^{3-\tau} & k\gg n^{\frac{\tau-2}{\tau-1}}, \tau>5/2\\
 k^{8-3\tau}n^{{\tau-3}} & n^{\frac{\tau-2}{\tau-1}}\ll k\ll\sqrt{n}, \tau<5/2,\\
 k^{\tau-2}n^{2-\tau} & \sqrt{n}\ll k\ll n^{1/(\tau-1)}/\log(n), \tau<5/2.
 \end{cases}
 \end{equation} 
 Thus, the closure coefficient undergoes a transition at $\tau=5/2$. This transition is caused by the typical structure of a triangle in the hyperbolic model. Whereas for $\tau>5/2$ a typical triangle from a vertex of degree $k$ contains two other vertices of constant degree, for $\tau<5/2$ it contains two other vertices of degree $n/k$ when $k$ is sufficiently large~\cite{stegehuis2019}.
 
 Furthermore, for $\tau$ close to 2 or 3, the slope of the closure coefficient in $k$ becomes small, making it almost $k$-independent for large $k$ and $\tau$ close to 2 or 3. 
 
 \paragraph{Relation with median closure.}
  The typical closure coefficient scales as~\eqref{eq:clhid} with high probability. Therefore, similarly as in the hidden-variable model, this suggests that the large-network behavior of the typical closure coefficient is similar to the median higher-order closure coefficient in the hyperbolic random graph. 
 Figure~\ref{fig:clakhyp} illustrates the behavior of the median closure coefficient in the hyperbolic random graph. Figure~\ref{fig:hishyp} shows the density of the closure coefficient in the hyperbolic model for large and small values of $k$. For both large and small values, the distribution of the closure coefficient is skewed, indicating a difference between the average and the median, or typical closure coefficient.


  \begin{figure*}[tb]
 	\centering
 	 	\subfloat[]{
 	 		\centering
 	 		\includegraphics[width=0.4\textwidth]{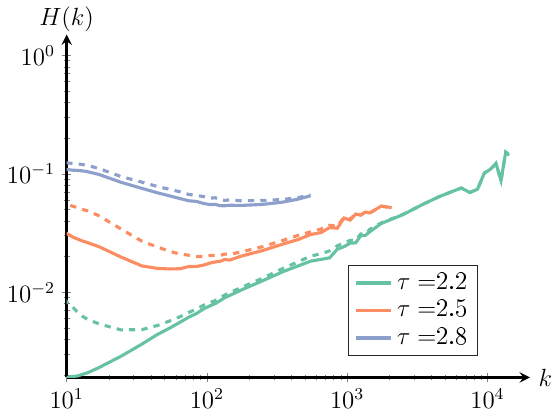}
 	 		\label{fig:cl2hyp}
 	 	}
 	 	\hspace{0.2cm}
 	\subfloat[]{
 		\centering
 		\includegraphics[width=0.4\textwidth]{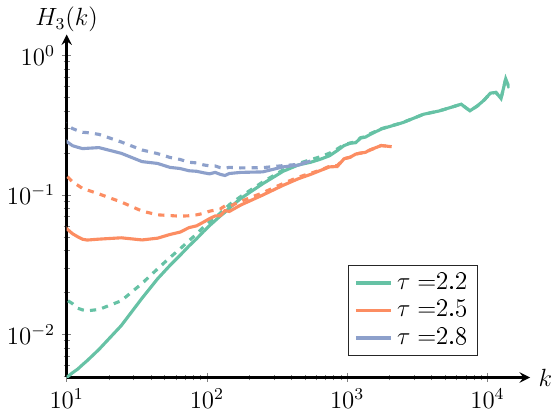}
 		\label{fig:cl3hyp}
 	}
 	\hspace{0.5cm}
 	\subfloat[]{
 		\centering
 		\includegraphics[width=0.4\textwidth]{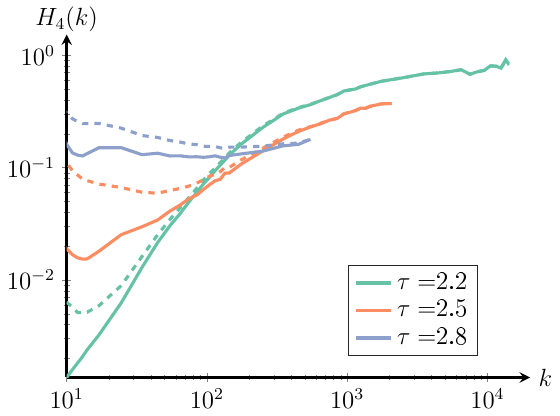}
 		\label{fig:cl4hyp}
 	}
 	\caption{The solid line plots the median value of the closure coefficient whereas the dashed line its average value, for the hyperbolic random graph with $n=10^5$ and various values of $\tau$ for a) $H(k)$, b) $H_3(k)$, c) $H_4(k)$.}
 	\label{fig:clhyp}
 \end{figure*}

\subsection{Higher-order closure}
We now perform an exploratory analysis of the higher-order closure coefficients of the hyperbolic random graph using similar methods as in Section~\ref{sec:hid}. Here it is convenient to define \emph{type} of a vertex as
\begin{equation}
w_i=\me^{(R-r_i)/2}.
\end{equation}
By~\cite{stegehuis2017b}, the type of a vertex is close to its degree. Therefore, we will first analyze the higher-order closure coefficient of a randomly chosen vertex with type $w$, which we denote by 
\begin{equation}\label{eq:closhighhat}
\hat{H}^{(t)}_l(w)=\frac{lK_{l+1}(V_w)}{W^{(l)}(V_h)},
\end{equation}
where $V_w$ is a randomly chosen vertex of type $w$. In Appendix~\ref{sec:typedegree}, we show that $\hat{H}^{(t)}_l(w)$ is close to ${H}^{(t)}_l(k)$ for $w=k$.

\paragraph{Cliques in the hyperbolic model with type-$w$ vertex.}
We first investigate the probability that a vertex of type $w$ forms a clique with $l\geq 3$ vertices of types $n^{\alpha}$.
In the hyperbolic random graph, the probability that vertex $i$ of type $n^\alpha$ connects to vertex $j$ of type $w$ satisfies~\cite{stegehuis2019}
\begin{equation}\label{eq:hypneibrprob}
\Prob{i\leftrightarrow j}\propto \min(wn^{\alpha-1},1).
\end{equation}
Thus, the probability that the type-$w$ vertex connects to the $l$ vertices of type $n^{\alpha}$ scales as $\min(wn^{\alpha-1},1)^{l}$. W.l.o.g., assume that the type-$w$ vertex is located at angle $\phi=0$. Then, the angle of a randomly chosen neighbor of the type-$w$ vertex of type proportional to $n^{\alpha}$ is uniformly distributed in an interval $[a,b]$, where $a\propto \min(wn^{\alpha-1},1)$ and $b\propto -\min(wn^{\alpha-1},1)$~\cite{stegehuis2019}.
These neighbors of the type-$w$ vertex form a clique if they are sufficiently close. Two vertices of types proportional to $n^{\alpha}$ are connected if their relative angle is at most $2\nu n^{2\alpha-1}$~\cite{bode2015}. Thus, for the $l$ vertices of type proportional to $n^\alpha$ to form a clique, their angles need to fall in the same interval of length proportional to $\min(n^{2\alpha-1},1)$.

The probability that $l$ vertices have angular coordinates in a given interval of width proportional to $\min(n^{2\alpha-1},1)$ conditionally on the fact that they are uniformly distributed in some interval $[a,b]$ of width proportional to $\min(wn^{\alpha-1},1)$ is
\begin{align}
\Prob{l \text{ neighbors of type-$w$ vertex form clique}} &\propto \left(\min\left(\frac{\min(n^{2\alpha-1},1)}{\min(wn^{\alpha-1},1)},1\right)\right)^{l-1}\nonumber\\
& =\min(n^\alpha\max(n^{\alpha-1},w^{-1}),1)^{l-1}.
\end{align}
Here the power $l-1$ comes from the fact that for $l$ vertices to fall in the same interval, we can first choose the position of the first vertex. This vertex defines the center of this interval, and the other $l-1$ vertices then have to be positioned in this interval to be sufficiently close to one another.

Combining this with~\eqref{eq:hypneibrprob} yields
\begin{align}\label{eq:pcliquehyp}
& \Prob{l+1\text{-clique with vertex of type }w \text{ and $l$ vertices of type } n^{\alpha}}\nonumber\\
& \propto \min(wn^{\alpha-1},1)^{l}\min(n^\alpha\max(n^{\alpha-1},w^{-1}),1)^{l-1},
\end{align}
 Since the type distribution of the hyperbolic random graph is also given by~\eqref{eq:deg_distr}~\cite[Lemma~1.3]{bode2015}, the number of vertices of weights proportional to $n^\alpha$ is with high probability proportional to $n^{1+\alpha(1-\tau)}$, as in the hidden-variable model. 
  Combining this with~\eqref{eq:pcliquehyp} yields that for $w\gg 1$
  \begin{align}
  \text{ \# $(l+1)$-cliques with }&V_h\text{ and $l$ vertices of types }n^{\alpha}\nonumber\\
  &\propto_{\sss{\prob}} n^{l+ \alpha l(1-\tau)}\min(wn^{\alpha-1},1)^{l} \min(n^\alpha\max(n^{\alpha-1},w^{-1}),1)^{l-1}.
  \end{align}
  Then,  similarly to~\eqref{eq:cliquemaxeq}, for $w\gg 1$,
\begin{align}\label{eq:optKuclique}
K_{l+1}(V_w)\propto_{\sss{\prob}}\max_{\alpha\in[0,1/(\tau-1)]} \  &  n^{l+ \alpha l(1-\tau)}\min(wn^{\alpha-1},1)^{l} \min(n^\alpha\max(n^{\alpha-1},w^{-1}),1)^{l-1},
\end{align}
if this equation has a unique optimizer.
The unique optimizer $\alpha^*$ is given by
\begin{equation}\label{eq:optalphKu}
n^{\alpha^*} = \begin{cases}
n^0 & \tau>3-\frac{1	}{l}\\
w & \tau<3-\frac{1}{l}, w\ll\sqrt{n}\\
n/w & 3-\frac{2}{l}<\tau<3-\frac{1}{l},w\gg \sqrt{n}\\
\sqrt{n} & \tau<3-\frac{2}{l},w\gg\sqrt{n}. 
\end{cases}
\end{equation}
Then, we obtain from~\eqref{eq:optKuclique}
\begin{equation}\label{eq:hypclique}
K_{l+1}(V_w)\propto_{\sss{\prob}}
\begin{cases}
w & \tau>3-\frac{1	}{l}, w\gg 1\\
w^{l(3-\tau)} & \tau<3-\frac{1}{l}, 1\ll w\ll\sqrt{n}\\
(n/w)^{l(3-\tau)-1}k&  3-\frac{2}{l}<\tau<3-\frac{1}{l},w\gg \sqrt{n}\\
n^{(3-\tau)l/2} & \tau<3-\frac{2}{l},w\gg\sqrt{n}. 
\end{cases}
\end{equation}

\paragraph{Wedges in the hyperbolic model with type-$w$ vertex.}
We now proceed to analyze the number of $l$-wedges attached to a randomly chosen vertex of type $w$.
By~\eqref{eq:hypneibrprob} and the fact that the vertex types are distributed as~\eqref{eq:deg_distr}, the number of neighbors of a vertex of type $w$ that have type proportional to $n^{\alpha}$ scales as $n^{\alpha(1-\tau)+1}\min(wn^{\alpha-1},1)$ with high probability. 
Furthermore, with high probability, a vertex of type $n^\alpha$ is part of $K_l(n^\alpha)$ cliques, where $K_l(n^\alpha)$ is as in~\eqref{eq:hypclique}. 

Thus, the number of $l$-wedges where the neighbor of the type-$w$ vertex has type proportional to $n^\alpha$ satisfies for $w\gg 1$
\begin{align}\label{eq:optWhyp}
&\text{\# $l$-wedges attached to $V_w$ where neighbor of $V_w$ has type $n^\alpha$}\nonumber\\
 &\propto_{\sss{\prob}} n^{\alpha(1-\tau)+1}\min(wn^{\alpha-1},1)K_l(n^\alpha),
\end{align}
with $K_l(n^\alpha)$ as in~\eqref{eq:hypclique}.
Similarly to~\eqref{eq:wedgemaxeq}, for $w\gg 1$,
\begin{equation}\label{eq:optWhyp2}
W^{(l)}(V_w)\propto_{\sss{\prob}}\max_{\alpha\in[0,1/(\tau-1)]} n^{\alpha(1-\tau)+1}\min(wn^{\alpha-1},1)K_l(n^\alpha),
\end{equation}
if this has a unique optimizer over $\alpha$. 
The unique optimizer $\alpha^*$ is given by
\begin{equation}
n^{\alpha^*} = \begin{cases}
n^{1/(\tau-1)} &\tau>3-\frac{2}{l} , w\ll n^{\frac{\tau-2}{\tau-1}}\\
n/w &  \tau>3-\frac{2}{l}, w\gg n^{\frac{\tau-2}{\tau-1}}\\
\sqrt{n} & \tau< 3-\frac{2}{l}.
\end{cases}
\end{equation}

Combining this with~\eqref{eq:optWhyp2} results in
\begin{equation}
W^{(l)}(V_w)\propto_{\sss{\prob}} \begin{cases}
 w^{\tau -2} n^{3-\tau }& \tau > 3-\frac{1}{l-1}, w\gg n^{\frac{\tau -2}{\tau -1}}\\
 w n^{\frac{3-\tau }{\tau -1}} &  \tau > 3-\frac{1}{l-1},1\ll  w\ll n^{\frac{\tau-2}{\tau -1}}\\
 w n^{\frac{1}{2} (l (3-\tau )-1)} &  \tau <3-\frac{1}{l-1},  1\ll w\ll \sqrt{n}\\
 n \left(\frac{n}{w}\right)^{l (3-\tau )-2}&  3-\frac{2}{l}<\tau <3-\frac{1}{l-1},
 w\gg\sqrt{n}\\
 n^{\frac{1}{2} l (3-\tau )} & \tau <3-\frac{2}{l}, w\gg\sqrt{n}.
\end{cases}
\end{equation}
Combining this with~\eqref{eq:closhighhat} yields for the higher-order closure coefficients in the hyperbolic random graph
\begin{equation}\label{eq:Hhighhyp}
\hat{H}_l^{(t)}(w)\propto_{\sss{\prob}} \begin{cases}
(w/n)^{3-\tau} & \tau<3-\frac{1}{l-1}, 1\ll w\ll\sqrt{n}\\
1 & \tau<3-\frac{1}{l-1}, w\gg\sqrt{n}\\
w^{l(3-\tau)-1}n^{\frac{\tau-3}{\tau-1}} &  3-\frac{1}{l-1}<\tau<3-\frac{1}{l}, 1\ll w\ll n^{\frac{\tau-2}{\tau-1}}\\
w^{2-\tau+l(3-\tau)}n^{3-\tau} & 3-\frac{1}{l-1}<\tau<3-\frac{1}{l},  n^{\frac{\tau-2}{\tau-1}}\ll w\ll \sqrt{n}\\
(w/n)^{4-\tau-l(3-\tau)}&  3-\frac{1}{l-1}<\tau<3-\frac{1}{l},  w\gg \sqrt{n}\\
n^{\frac{\tau-3}{\tau-1}} & \tau>3-\frac{1}{l}, 1\ll w\ll n^{\frac{\tau-2}{\tau-1}}\\
(w/\sqrt{n})^{l(3-\tau)-1} & \tau>3-\frac{1}{l}, w\gg n^{\frac{\tau-2}{\tau-1}}.
\end{cases}
\end{equation}
In Appendix~\ref{sec:typedegree}, we show that $H^{(t)}(k)$ follows the same scaling.

In particular,~\eqref{eq:Hhighhyp} shows that for any value of $\tau\in(2,3)$ there exists an $l^*\geq 2$ such that $H_l^{(t)}(k)$ is of constant order of magnitude for all $l\geq l^*$ and $k\gg\sqrt{n}$. 
This implies that sufficiently large cliques linked to a vertex of degree $k\gg\sqrt{n}$ are likely to be part of larger cliques. This also occurs in the hidden-variable model (see Eq~\eqref{eq:chighhid}). However, for the hidden-variable model, this already happens for cliques of size 3. In the hyperbolic random graph on the other hand, $H_l^{(t)}$ scales as a constant when $\tau<3-1/(l-1)$, or when $l>1+1/(3-\tau)$. Thus, cliques of size $1+1/(3-\tau)>3$ for $\tau\in(2,3)$ only start to be likely to be part of a larger clique.   
Eq.~\eqref{eq:Hhighhyp} also shows that $H_l(k)$ is always non-decreasing in $k$. 

\paragraph{Relation with median closure.}
Figures~\ref{fig:cl3hyp} and~\ref{fig:cl4hyp} show the behavior of the median higher-order closure coefficients in the hyperbolic random graph. 
These figures show that the slope of $H_4^{(t)}(k)$ becomes very small when $k$ gets large, as predicted by~\eqref{eq:Hhighhyp}. 

 Furthermore, Figure~\ref{fig:clhyp} shows that the median or typical values of the higher-order closure coefficients indeed behave significantly different from their average values, as predicted in Section~\ref{sec:typ}. Especially for small values of $k$, the average closure coefficients are much higher than their median values. This indicates the presence of a few low-degree vertices with extremely large values of their higher-order closure coefficients that dominate the mean value of $H_l(k)$. The typical behavior of $H_l(k)$ is illustrated in Figure~\ref{fig:clhyp} by the median value of the local closure coefficient of vertices of degree $k$, which is less affected by such outliers. Therefore, analyzing the typical behavior of the local closure coefficient of a vertex of degree $k$ results in a lower value than its average.

\section{Conclusion}
In this paper, we have explored the behavior of the closure coefficient in two random graph models that create simple, scale-free networks: the hidden-variable model and the hyperbolic random graph. 
For both models we have obtained asymptotic expressions of the closure coefficient as well as all higher-order closure coefficients. 

The behavior of the closure coefficients in these random graph models is significantly different from its behavior in the configuration model~\cite{yin2019}, a random graph model that creates multigraphs. This shows that the degree-correlations arising from the simplicity constraint in the models we have analyzed has a significant impact on this new measure for clustering. 

The difference between the hyperbolic random graph and the hidden-variable model in terms of the behavior of their closure coefficients indicates that the presence of geometry mostly influences the presence of small cliques. In the hyperbolic model, these are more likely to be formed between low-degree vertices, whereas in the hidden-variable model they are mostly formed with vertices of degree at least $\sqrt{n}$. Large cliques are usually formed in the core of vertices of degree at least $\sqrt{n}$ in both models.  
%

In this paper, we performed an exploratory analysis of the scaling of the closure coefficient in terms of the network size $n$. It would be interesting to obtain more precise results on its behavior. We conjecture that the rescaled closure coefficient in the hidden-variable model converges in probability to a constant. Finding this constant can probably be achieved by using integral equations as in~\cite{hofstad2017d}. For the hyperbolic model, we believe that for $k$ sufficiently large, the closure coefficients will also converge in probability to a constant. However, deriving this constant will be more difficult than for the hidden-variable model. The behavior of the local closure coefficient for small $k$ is another interesting open question.  

Furthermore, in this paper we investigated the behavior of the closure coefficient of a randomly chosen vertex of degree $k$, which behaves similarly as the median closure coefficient. It would be interesting to also see analytic results for its average value over all vertices of degree $k$. Our simulations suggest that for small values of $k$, these two quantities may behave significantly different, but they seem close for larger values of $k$. Finding the value of $k$ where both statistics behave similarly is therefore an interesting question for further research.

Another interesting direction for future work is investigating the typical motif conductance of vertices of degree $k$, which has recently been shown to be related to higher-order closure coefficients~\cite{yin2019}. In particular, our results in~\eqref{eq:chighhid} then and~\eqref{eq:Hhighhyp} give lower bounds on the $l$-clique conductance of neighborhoods of vertices of degree $k$. A low motif conductance of a vertex indicates it may be a good starting point for a community detection algorithm. Therefore, pursuing this line of research to investigate the motif conductance could lead to better understanding of community detection algorithms.


\appendix
\section{From weights to degrees}\label{sec:weightdegree}
Let $f(h,n)$ denote the scaling of the typical closure coefficient of a randomly chosen of weight $h$ as predicted by~\eqref{eq:chighhidweight}. That is,
\begin{equation}\label{eq:fhn}
f(h,n)=\begin{cases}
h^{l-1}n^{-(l-1)/2} & 1\ll h\ll \sqrt{n}\\
1 & h\gg\sqrt{n}.
\end{cases}
\end{equation}
We now show that ${H}_l^{(t)}(k)$ also scales as $f(h,n)$. Conditionally on the weights, the degree of a vertex $v$, $D_v$, is the sum of $n-1$ independent indicators indicating the presence of an edge between vertex $v$ and the other vertices. Furthermore, the connection probability~\eqref{eq:conprobhvm} ensures that $\Exp{D_v\mid h_v}=h_v$. Thus, by the Chernoff bound, for any $\delta>1$,
\begin{equation}\label{eq:wvdvbound1}
\Prob{D_v\geq h_v(1+\delta)\mid h_v}\leq \exp(-\delta h_v/3).
\end{equation}
Similarly,
\begin{equation}
\Prob{D_v\leq h_v(1-\delta)\mid h_v}\leq \exp(-\delta h_v/2).
\end{equation}
Furthermore,
\begin{align}
\Prob{h_v<(1-\varepsilon)k\mid D_v=k} & = \frac{\Prob{D_v=k,h_v<(1-\varepsilon)k}}{\Prob{D_v=k}}\nonumber\\
& = \frac{\Prob{D_v=k\mid h_v<(1-\varepsilon)k}\Prob{h_v<(1-\varepsilon)k}}{\Prob{D_v=k}}\nonumber\\
& \leq  \frac{\Prob{D_v\geq k\mid h_v<(1-\varepsilon)k}\Prob{h_v<(1-\varepsilon)k}}{\Prob{D_v=k}}\nonumber\\
& \leq  \frac{\Prob{D_v\geq k\mid h_v=(1-\varepsilon)k}\Prob{h_v<(1-\varepsilon)k}}{\Prob{D_v=k}},
\end{align}
where the last line follows because the probability that a vertex with weight $h_1$ has degree
at least $k$ is larger than the probability that a vertex of weight $h_2$ has degree at least $k$ when $h_1>h_2$.
Since $\Prob{D_v=k}\propto k^{-\tau}$ and $\Prob{h_v<(1-\varepsilon)k}\propto 1-k^{1-\tau}$ by~\eqref{eq:deg_distr}, 
\begin{align}
\Prob{h_v<(1-\varepsilon)k\mid D_v=k} & \propto k^\tau(1-k^{1-\tau})\Prob{D_v\geq k\mid h_v=(1-\varepsilon)k}\nonumber\\
& \propto k^\tau\Prob{D_v\geq \tilde{k}(1+\varepsilon/(1-\varepsilon))\mid h_v=\tilde{k}}\nonumber\\
& =O\left(k^\tau \exp\left(-\frac{k\varepsilon}{3}\right)\right),
\end{align}
where $\tilde{k}=k(1-\varepsilon)$ and where we used~\eqref{eq:wvdvbound1}.
Similarly,
\begin{align}
\Prob{h_v>(1+\varepsilon)k\mid D_v=k} & =O\left(k^\tau \exp\left(-\frac{\varepsilon k}{2}\right)\right).
\end{align}
Therefore, for any vertex $v$ with $h_v\gg 1$, $h_v=D_v(1+\op(1))$. Furthermore,~\eqref{eq:fhn} shows that $f(k(1+o(1)),n)\propto f(k,n)$ for $k\gg 1$. 
As the higher order closure coefficient of a vertex of weight $h$ is proportional to $f(h,n)$ with high probability and the vertex has degree $h(1+\op(1))$, this also implies that                                                                                                                                                                                                                                                                                                                                                                                                                                                                                                                                                                                                                                                                                                                                                                                                                                                                                                                                                                                                                                                                                                                                                                                                                                                                                                                                                                                                                                                                                                                                                                                                                                                                                                                                                                                                                                                                                                                                                                                                                                                                                                                                                                                                                                                                                                                                                                                                                                                                                                                                                                                                                                                                                                                                                                                                                                                                                                                                                                                                                                                                                                                                                                                                                                                                                                                                                 
\begin{equation}
{H}_l^{(t)}(k)\propto_{\sss{\prob}}\begin{cases}
k^{l-1}n^{-(l-1)/2} &1\ll  k\ll \sqrt{n}\\
1 & k\gg\sqrt{n}.
\end{cases}
\end{equation}


\section{From types to degrees}\label{sec:typedegree}
Showing that $\hat{H}^{(t)}(w)$ has the same scaling as $H^{(t)}(k)$ in the hyperbolic model follows the same lines as the proof in Appendix~\ref{sec:typedegree} for the hidden-variable model. Again, let $\tilde{f}(w,n)$ denote the scaling of $\hat{H}^{(t)}(w)$, so that 
\begin{equation}\label{eq:fwn}
\tilde{f}(w,n) \begin{cases}
(w/n)^{3-\tau} & \tau<3-\frac{1}{l-1}, w\ll\sqrt{n}\\
1 & \tau<3-\frac{1}{l-1}, w\gg\sqrt{n}\\
w^{l(3-\tau)-1}n^{\frac{\tau-3}{\tau-1}} &  3-\frac{1}{l-1}<\tau<3-\frac{1}{l}, w\ll n^{\frac{\tau-2}{\tau-1}}\\
w^{2-\tau+l(3-\tau)}n^{3-\tau} & 3-\frac{1}{l-1}<\tau<3-\frac{1}{l},  n^{\frac{\tau-2}{\tau-1}}\ll w\ll \sqrt{n}\\
(w/n)^{4-\tau-l(3-\tau)}&  3-\frac{1}{l-1}<\tau<3-\frac{1}{l},  w\gg \sqrt{n}\\
n^{\frac{\tau-3}{\tau-1}} & \tau>3-\frac{1}{l}, w\ll n^{\frac{\tau-2}{\tau-1}}\\
(w/\sqrt{n})^{l(3-\tau)-1} & \tau>3-\frac{1}{l}, w\gg n^{\frac{\tau-2}{\tau-1}}
\end{cases}
\end{equation}

By,~\cite[Eq~(112)]{stegehuis2017b}, for any $\varepsilon>0$ and $D_v\gg 1$,
\begin{equation}
\lim_{n\to\infty}\Prob{w_v>(1+\varepsilon)\frac{\pi(\tau-2)}{2\nu (\tau-1)}D_v} = 0
\end{equation}
and
\begin{equation}
\lim_{n\to\infty}\Prob{w_v<(1-\varepsilon)\frac{\pi(\tau-2)}{2\nu (\tau-1)}D_v} = 0.
\end{equation}
Therefore, for any vertex $v$ with $w_v\gg 1$, $w_v=\frac{\pi(\tau-2)}{2\nu (\tau-1)}D_v(1+\op(1))$. Furthermore, it can be seen from~\eqref{eq:fwn} that $\tilde{f}(w,n)\propto \tilde{f}(w(1+o(1)),n)$. 
This shows that $H^{(t)}(k)\propto_{\sss{\prob}} \tilde{f}(k,n)$.

\section{$c(k)$ hidden-variable model}\label{sec:ckhvm}
The proof of~\eqref{eq:ckhvm} for the hidden-variable model follows the same lines of the proof of~\cite[Theorem~1]{hofstad2017c}
where the behavior for $c(k)$ in the erased configuration model is analyzed, as also stated on~\cite[remark on page 7]{hofstad2017c}. This proof approximates the connection probabilities of vertices $i$ and $j$ in the erased configuration model by $1-\me^{-D_iD_j/(\mu n)}$, where $\mu$ denotes the average degree. In the hidden-variable model, the connection probability equals $\min(h_ih_j/(\mean{h}n),1)$. Thus, we can follow the lines of the proof of Theorem 1, where in the computation of the constant $A$ in (3.16), we replace $1-\me^{-x}$ by $\min(x,1)$. This gives
\begin{equation}
A=\int_{0}^{\infty}t^{1-\tau}\min(t,1)\dd t = \frac{1}{(3-\tau)(\tau-2)}.
\end{equation}

Then,~\cite[(3.2)]{hofstad2017c} becomes for $ n^{(\tau-2)/(\tau-1)}\ll k\ll \sqrt{n}$
\begin{align}
\frac{c(k)}{n^{2-\tau}\log(n/k^2)}& 
\plim 
  C^2\mean{h}^{-\tau}\frac{1}{(3-\tau)(\tau-2)}.
\end{align}
Similarly,~\cite[(3.3)]{hofstad2017c} becomes for $\sqrt{n}\ll k\ll n^{1/(\tau-1)}/\log(n)$, 
\begin{align}
\frac{{c(k)}}{n^{5-2\tau}k^{2\tau-6}}& \plim C^2\mean{h}^{3-2\tau}\left(\frac{1}{(3-\tau)(\tau-2)}\right)^2.
\end{align}

\end{document}